\documentclass{gmee2012}
\usepackage{cite}      
\usepackage{graphicx}
\usepackage{epstopdf}
\usepackage{psfrag} 
\usepackage{amsmath}   
\DeclareMathOperator{\varRe}{Re}
\DeclareMathOperator{\varIm}{Im}

\usepackage[load-configurations=binary]{siunitx}
\usepackage{booktabs}
\usepackage{fixltx2e}
\usepackage{paralist}
\usepackage{marvosym}

\newcommand{\ju}{\ensuremath{\mathrm{j}}}

\begin{document}
%
\title{\LARGE{Experiences with a two terminal-pair \\ digital impedance bridge}}

\author{Luca Callegaro, Vincenzo D'Elia, Marian Kampik, \\ Dan Bee Kim, Massimo Ortolano, and Faranak Pourdanesh
\thanks{Luca Callegaro, Vincenzo D'Elia and Faranak Pourdanesh are with the Electromagnetism Division of the Istituto Nazionale di Ricerca Metrologica (INRIM), strada delle Cacce 91, 10135 Torino, Italy. E-mail: \texttt{l.callegaro@inrim.it}.} 
\thanks{Marian Kampik is with the Silesian University of Technology, Poland.}
\thanks{Dan Bee Kim is with KRISS - Korea Research Institute of Standards and Science, Daejeon, Korea.}
\thanks{Massimo Ortolano is with the Politecnico di Torino, Torino, Italy.}
\thanks{Manuscript received \today.}
}
%

\maketitle

\begin{abstract}
This paper describes the realization of a two terminal-pair digital impedance bridge and the test measurements performed with it. The bridge, with a very simple architecture, is based on a commercial two-channel digital signal synthesizer and a synchronous detector. The bridge can perform comparisons between impedances having arbitrary phase and magnitude ratio: its balance is achieved automatically in less than a minute. $R$-$C$ comparisons with calibrated standards, at \si{\kilo\hertz} frequency and \SI{100}{\kilo\ohm} magnitude level, give ratio errors of the order of \num{E-6}, with potential for further improvements.
\end{abstract}

%

\section{Introduction}
We implemented a coaxial voltage ratio bridge to perform comparisons in the audio frequency range of two terminal-pair impedance standards of arbitrary magnitude ratio and phase difference. The bridge is \emph{digital}: its main component is a two-channel digital signal source. 

\begin{figure}[htb]
	\centering
	\def\svgwidth{0.4\linewidth}	
	\input{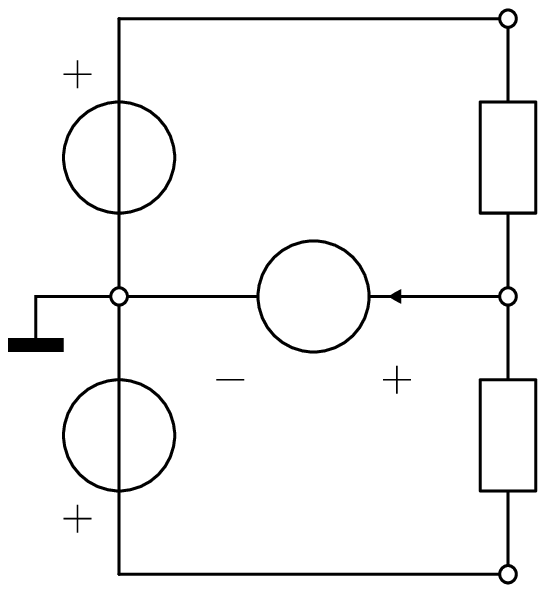}
 	\caption{Schematic diagram of the digital bridge, see text for details. \label{fig:principle}}%
\end{figure}

The bridge, introduced in~\cite{Callegaro2014_CPEMDigitalBridge}, is here described in full detail, together with test measurements and an expression of the measurement uncertainty.

\subsection{Bridge principle}
The schematic diagram of the bridge, well known in the literature (see~\cite[Ch.~5]{CallegaroBook} and references therein; \cite{Kim2012,Lan2012}), is given in Fig.~\ref{fig:principle}. The source output channel $E_1$ drives the impedance $Z_\textup{A}$ (admittance $Y_\textup{A}=1/Z_\textup{A}$); channel $E_2$, the impedance $Z_\textup{B}$ (admittance $Y_\textup{B}=1/Z_\textup{B}$). $Z_\textup{A}$ and $Z_\textup{B}$ are in series and the null detector D senses the common voltage at the low terminals of $Z_\textup{A}$ and $Z_\textup{B}$. The bridge balance condition $V_\textup{D} = 0, I_\textup{D} = 0$ is achieved by adjusting amplitude and phase of one of the two channels. At equilibrium, $E_1 Y_\textup{A} + E_2 Y_\textup{B} = 0$: this implies that the complex impedance ratio $W = Z_\textup{A}/Z_\textup{B}$ is given by
\begin{equation}
\label{eq:basicmodel}
  W = \frac{Z_\textup{A}}{Z_\textup{B}} = -\frac{E_1}{E_2}\,.
\end{equation}
The pair $E_1,E_2$ constitutes the bridge reading.
\subsection{Measurement model}
\label{sec:measmodel}
\begin{figure}[htb]
	\centering
	\def\svgwidth{0.7\linewidth}	
	\input{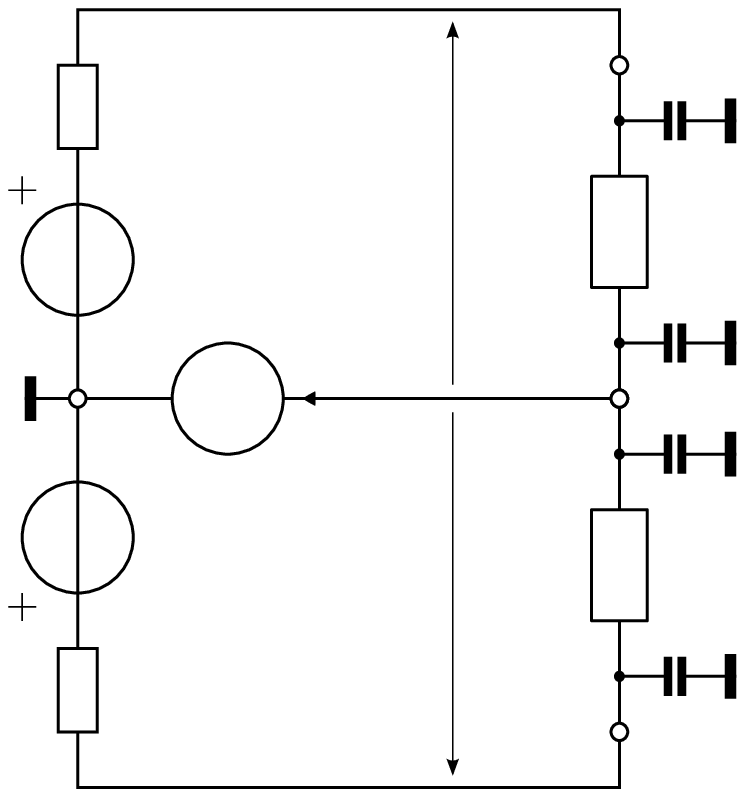}
 	\caption{Circuit model employed in the determination of the measurement model of Sec.~\ref{sec:measmodel} (see text for details); thick segments (\Flatsteel) represent connections to the shield. \label{fig:model}}%
\end{figure}

The schematic diagram of Fig.~\ref{fig:principle} represents an idealized bridge. Fig.~\ref{fig:model}, instead, shows a circuit model which takes into account the source output impedances and the stray capacitances of the impedance standards in two-terminal pair definition. 
Assuming that the impedances under comparison are defined at the end of the connecting cables, they can be modeled as two-port $\Pi$ networks. Each $\Pi$ network comprises the high-to-low transadmittance $Y_\textup{X}$ (where $\textup{X} = \textup{A}, \textup{B}$), the high-to-shield admittance $y_\textup{HX}$ and the low-to-shield admittance $y_\textup{LX}$. Typically, $y_\textup{HX}$ and $y_\textup{LX}$ can be regarded as purely capacitive, with an equivalent capacitance of the order of \SI{100}{\pico\farad}.

Each channel $k = 1,2$ can be modeled with a Th\'{e}venin equivalent circuit composed of an ideal voltage source $E_{k\textup{X}}$ in series with an output impedance $z_k$. At equilibrium, when the source $k$ is connected to the impedance $Y_\textup{X}$, the channel output voltage $V_{k\textup{X}}$ is 
\begin{equation}
V_{k\textup{X}} = \frac{1}{1+z_k \left(Y_\textup{X} + y_\textup{HX} \right)} E_{k\textup{X}}.
\end{equation}

It is well known that exchanging the standards under comparison in the bridge arms can correct some of the systematic errors. We call \emph{forward}~(F) the configuration where $Y_\textup{A}$ is connected to source channel $1$ and $Y_\textup{B}$ to channel $2$; \emph{reverse}~(R) the configuration where $Y_\textup{A}$ is connected to channel $2$ and $Y_\textup{B}$ to channel $1$. The equilibrium conditions for the two configurations can be written as
\begin{align}
W &= -\frac{V_{1\textup{A}}}{V_{2\textup{B}}} = -\frac{E_{1\textup{A}}}{E_{2\textup{B}}} \left(\frac{1+z_2 \left(Y_\textup{B} + y_\textup{HB} \right) }{1+z_1 \left(Y_\textup{A} + y_\textup{HA} \right)} \right) \quad \textup{(F)}, \nonumber \\
W &= -\frac{V_{2\textup{A}}}{V_{1\textup{B}}} = -\frac{E_{2\textup{A}}}{E_{1\textup{B}}} \left(\frac{1+z_1 \left(Y_\textup{B} + y_\textup{HB} \right) }{1+z_2 \left(Y_\textup{A} + y_\textup{HA} \right)} \right) \quad \textup{(R)}. 
\end{align}
Because of source imperfection, the actual ratio $E_1/E_2$ deviates from the reading $E_1^{(\textup{r})}/E_2^{(\textup{r})}$. We model this deviation with a complex gain tracking error $g$, dependent on the channel setting: 
\begin{equation}
\frac{E_1^{(\textup{r})}}{E_2^{(\textup{r})}} = (1+g) \frac{E_1}{E_2}\,.
\end{equation}
By taking the geometric average of the forward and the reverse bridge readings, the measurement model can be written as
\begin{multline}
\label{eq:fullmodel}
W = \bigg[\frac{1+g_\textup{R}}{1+g_\textup{F}}\,\frac{E_{1\textup{A}}^{(\textup{r})}}{E_{2\textup{B}}^{(\textup{r})}}\,\frac{E_{2\textup{A}}^{(\textup{r})}}{E_{1\textup{B}}^{(\textup{r})}}	\\ \times \frac{1 + z_2 \left(Y_\textup{B} + y_\textup{HB} \right)}{1 + z_1 \left(Y_\textup{A} + y_\textup{HA} \right)}\,\frac{1 + z_1 \left(Y_\textup{B} + y_\textup{HB} \right)}{1 + z_2 \left(Y_\textup{A} + y_\textup{HA} \right)}\bigg]^{\frac{1}{2}}\,,
\end{multline}
where $g_\textup{F}$ and $g_\textup{R}$ are respectively the forward and reverse gain tracking errors. Eq.~\eqref{eq:fullmodel} actually yields two values: the choice of the proper branch for the square root should be made according to the nominal value of $W$.   

Under the assumptions that $|g_\textup{F}|, |g_\textup{R}| \ll 1$ and that all terms $|z(Y+y_\textup{H})| \ll 1$, Eq.~\eqref{eq:fullmodel} can be linearized as 
\begin{equation}
\label{eq:fullmodel_linear}
W = W^{(\textup{r})}(1+\epsilon_W)\,,
\end{equation}
where 
\begin{equation}
\label{eq:ratio_reading}
W^{(\textup{r})} = \left[ \frac{E_{1\textup{A}}^{(\textup{r})}}{E_{2\textup{B}}^{(\textup{r})}} \frac{E_{2\textup{A}}^{(\textup{r})}}{E_{1\textup{B}}^{(\textup{r})}}	\right]^{\frac{1}{2}}
\end{equation}
is the ratio reading and
\begin{equation}
\label{eq:epsilonW}
\epsilon_W = - \frac{1}{2} \Delta g_\textup{FR}  + \frac{1}{2} \left( z_1 + z_2 \right)\big[( Y_\textup{B} + y_\textup{HB}) - (Y_\textup{A} + y_\textup{HA})\big]\,,
\end{equation}
with $\Delta g_\textup{FR} = g_\textup{F} - g_\textup{R}$, is a correction term which accounts for the bridge nonidealities.

Eq.~\eqref{eq:fullmodel_linear} shows, as expected, that even a significant but setting-independent gain tracking error $g$ is compensated by averaging the two readings, whereas the error due to the output impedance is in general not compensated, even in 1:1 comparisons, because of the presence of the $y_\textup{HX}$ terms. 
\section{Implementation}
\begin{figure}[htb]
	\centering
	\def\svgwidth{0.6\linewidth}	
	\input{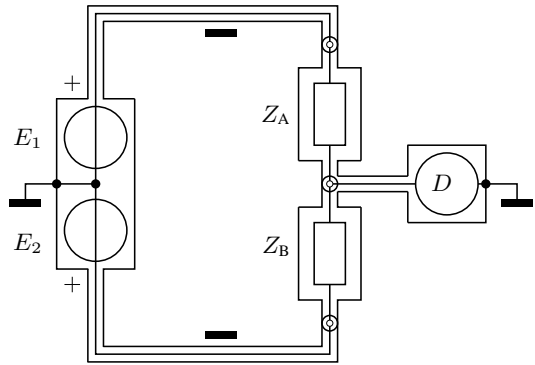}
 	\caption{Coaxial schematic diagram of the digital bridge (see text for details). The black rectangles identify coaxial equalizers. \label{fig:coaxscheme}}%
\end{figure}

\begin{figure}[htb]
	\centering
	\includegraphics[width=\linewidth]{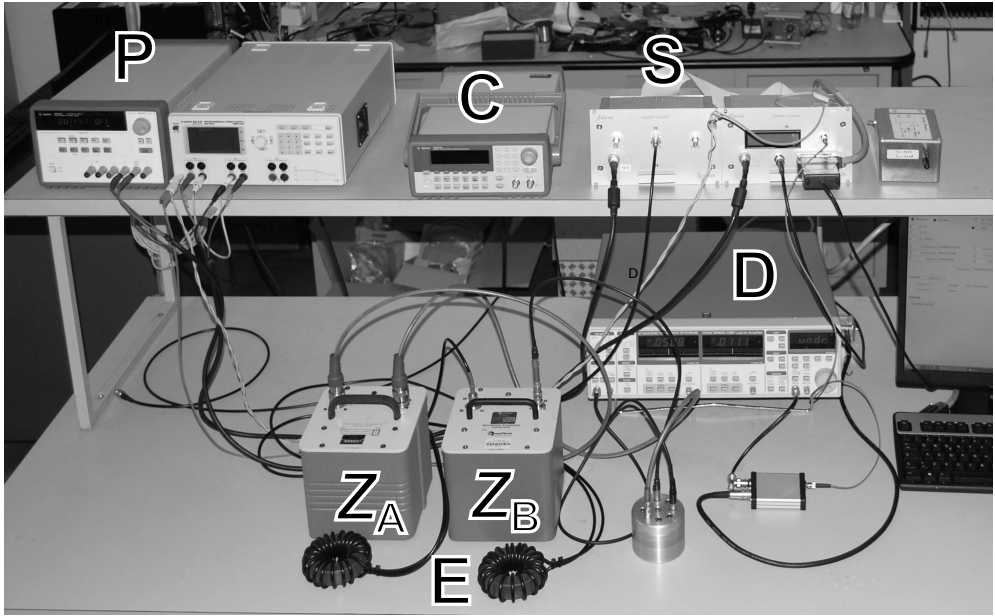}
	 \caption{Picture of the digital bridge showing the impedances $Z_\textup{A}$ and $Z_\textup{B}$ under comparison, the source S (energized by power supplies P, one for the analog and one for the digital part), the detector D and the coaxial equalizers E. A \SI{10}{\mega\hertz} clock source C is also shown.}%
	\label{fig:photo}
\end{figure}

A coaxial schematic diagram of the bridge is given in Fig.~\ref{fig:coaxscheme}; Fig.~\ref{fig:photo} shows a picture of the assembly.
The devices employed in this realization are:
\begin{itemize}
\item[Source (S).] Aivon Oy DualDAC (2 channels, \SI{16}{bit} resolution, up to \SI{5}{\mega S\per\second} maximum sampling rate, $2^{14}$ maximum sample buffer size; the digital part is optically isolated from the analog one). 
\item[Detector (D).] Stanford Research mod.~830 lock-in amplifier; an optical output from the source provide the reference signal.
\item[Equalizers (E).] Coaxial equalizers on nanocrystalline ferromagnetic cores.
\end{itemize} 

Amplitude and phase of each channel are adjusted by recalculating and uploading new waveform samples. Each sample code is chosen to minimize the quantization error. More refined synthesis strategies can be implemented to improve the resolution~\cite{Kampik:2014a}. The source implements a double buffer, which allows continuous output even during the upload of a new sample set. The quantities $E_{1\textup{X}}^{(\textup{r})}$, $E_{2\textup{X}}^{(\textup{r})}$ which appear in~\eqref{eq:ratio_reading} are calculated from the Fourier expansions of the quantized waveforms.

The control program implements a simple balancing strategy~\cite{Callegaro2005} which allows to reach a residual voltage $V_\textup{D}$ in the \SI{100}{\nano\volt} range in less than a minute.

\section{Experimental}
\subsection{Some properties of the source employed}
\label{sec:source_specs}
In model~\eqref{eq:fullmodel_linear}--\eqref{eq:epsilonW} the parameters related to the source are the source output impedances $z_1$, $z_2$ and $\Delta g_{\textup{FR}}$ which takes into account source nonidealities.

The impedances $z_1$ and $z_2$ were measured with an LCR meter Agilent mod.\ 4284A; for frequency $f$ up to about \SI{20}{\kilo\hertz}, the output impedance $z_k$ can be modelled with a resistance $r_k$ in series with an inductance $l_k$, $z_k = r_k + \ju 2\pi f l_k$, where $r_k = \SI{100(50)}{\milli\ohm}$ and $l_k = \SI{4(1)}{\micro\henry}$.  

The term $\Delta g_{\textup{FR}}$ has undergone a preliminary evaluation~\cite{Kampik:2014b} for $W$ ratios close to $-1$. The span of $|\Delta g_{\textup{FR}}|$ is less than \num{2e-6} for $W$ spanning a range of \num{2e-4} about $-1$. $\Delta g_{\textup{FR}}$ becomes more significant for values of $|W|$ far from unity; however, a full characterization of this parameter has not yet been completed.

Other nonidealities not considered in the model of Sec.~\ref{sec:measmodel} were evaluated and found negligible. The relative stability of $E_1/E_2$ over time of the source employed was tested with the bridge itself, by substituting the impedance standards with an inductive voltage divider (which has a negligible ratio drift). Results are reported in~\cite{Nissila2014}; the Allan deviation of the amplitude ratio at \SI{1}{\kilo\hertz} is \SI{10}{\nano\volt\per\volt} over \SI{30}{\minute}; phase difference fluctuations are dominated by flicker noise beyond \SI{100}{\second}, with an Allan deviation of \SI{40}{\nano\radian}. The crosstalk between the channels is lower than \SI{-125}{\deci\bel} up to \SI{16}{\kilo\hertz}.

\subsection{Impedance measurements}
The bridge was tested with the impedance standards listed in Tab.~\ref{tab:standards}, calibrated as two terminal-pair standards (at the end of the connecting cables).
\begin{table}[tb]
\centering
\caption{Standards employed during the measurements. The asterisk * denotes the second of two standards of the same model.}
\begin{tabular}{rl}
\toprule
Label & Description \\
\midrule
\SI{1}{\nano\farad}, *\SI{1}{\nano\farad} & General Radio mod. 1404-C, sealed N\textsubscript{2} \\ 
\SI{10}{\nano\farad}, *\SI{10}{\nano\farad} & Custom realization, C0G solid dielectric \cite{Callegaro2005b} \\ 
\SI{100}{\kilo\ohm} & Agilent 42039A  \\ 
\SI{10}{\kilo\ohm} & Agilent 42038A \\
\bottomrule
\end{tabular}
\label{tab:standards}
\end{table}

Tab.~\ref{tab:results} reports the measurement results. For each comparison, the reported values are: 
\begin{itemize}
\item The types and the nominal values of the impedances $Z_\textup{A}$ and $Z_\textup{B}$;
\item The measurement frequency $f$ chosen to have an angular frequency close to a decadic value;
\item The real and the imaginary parts of $W$ as computed from the measurement model (the operators $\varRe$ and $\varIm$ denote the real and the imaginary parts, respectively);
\item A reference ratio $W^\textup{ref} = Z_\textup{A}^\textup{ref} / Z_\textup{B}^\textup{ref}$ calculated from values $Z_\textup{A}^\textup{ref}$ and $Z_\textup{B}^\textup{ref}$ obtained by independent two-terminal pair calibrations traceable to the Italian national standards of capacitance and resistance;
\item The real and the imaginary parts of the deviation $\delta = W-W^\textup{ref}$ of the bridge measurement from the reference ratio.
\end{itemize}

It is worth pointing out the meaning of the components of $W$:
\begin{itemize}
\item in $C$-$C$ comparisons, $\varRe W$ is related to the capacitance ratio, while $\varIm W$ is related to the difference of the phase angles;
\item in $R$-$C$ comparisons, $\varIm W$ is related to the principal parameter of the impedances (the resistance and the capacitance), whereas $\varRe W$ is related to the secondary parameter (i.e., the resistor time constant and the capacitor phase angle); the fact that $\varRe\delta > \varIm\delta$ can be possibly due to the mediocre knowledge of these secondary parameters, for which INRIM does not have primary national standards. 
\end{itemize}

\section{Uncertainty}
Since the measurement model~\eqref{eq:fullmodel_linear}--\eqref{eq:epsilonW} is a complex-valued function of complex-valued input quantities, an expression of the bridge measurement uncertainty has to be carried out in the context of the Supplement 2 of the \emph{Guide to the expression of uncertainty in measurement}~\cite{GUMS2}. The calculations were performed with the \texttt{Metas.UncLib}~\cite{Zeier2012} software package.

An example of uncertainty budget is reported in Tab~\ref{tab:uncbug} for a comparison between a \SI{100}{\kilo\ohm} resistor and \SI{1}{\nano\farad} capacitor at the frequency of \SI{1592.36}{\hertz} (see also row 7 of Tab.~\ref{tab:results}).

Some notes about the evaluation of the uncertainties of the model input quantities:
\begin{itemize}
\item The measurements of Tab.~\ref{tab:results} and the uncertainty budget of Tab.~\ref{tab:uncbug} correspond to $|W|\approx 1$, for which we have a characterization of $\Delta g_\textup{FR}$. We assigned $\Delta g_\textup{FR}=0$ with an uncertainty compatible with the source specifications given in Sec.~\ref{sec:source_specs}.
\item $Y_\textup{A}$ and $Y_\textup{B}$ are known from their nominal values, with negligible uncertainty for what concerns the correction term $\epsilon_W$ given by~\eqref{eq:epsilonW};
\item $y_\textup{HA}$ and $y_\textup{HB}$ include also the connections, and are considered as pure capacitances, $y_\textup{HX} = \ju 2\pi f c_\textup{HX}$; an uncertainty is included to take into account variations in cable lengths and differences between models;
\item The uncertainty of $W^{(\textup{r})}$ is the type A uncertainty related to the measurement repeatability.
\end{itemize}

\begin{table*}[tb]
\centering
\caption{Results of comparisons performed with the bridge.}
\begin{tabular}{rrS[table-format=5.2]S[table-format=1.7e+1,table-align-exponent=false,table-column-width=18mm]S[table-format=+1.7e+1,table-align-exponent=false,table-column-width=18mm]S[table-format=+3.1]S[table-format=+2.1]}
\toprule
{$Z_\textup{A}$} & {$Z_\textup{B}$} & {$f/\si{\hertz}$} & {$\varRe W$} & {$\varIm W$} & {$\varRe\delta$} & {$\varIm\delta$} \\ 
&&&&& $\times \num{E6}$ & $\times \num{E6}$ \\
\midrule
\SI{1}{\nano\farad} & *\SI{1}{\nano\farad} & 159.24  		& 1.0003240 	& 1.83E-6		& 0.3	 & 0.1\\ 
\SI{1}{\nano\farad} & *\SI{1}{\nano\farad} & 1592.36  	& 1.0003226 	& 1.36E-6 		& -0.5 	 &-0.1\\ 
\SI{1}{\nano\farad} & *\SI{1}{\nano\farad} & 15873.02  	& 1.0003120 	& 2.01E-5		& -15.4  & 17 \\ 
\SI{10}{\nano\farad} & *\SI{10}{\nano\farad} & 159.24  	& 0.9999200 	& 5.23E-7		& -2.0 	& 0.0 \\ 
\SI{10}{\nano\farad} & *\SI{10}{\nano\farad} & 1592.36  	& 0.9999226 	& -3.18E-7	    & -1.4 & 0.6 \\ 
\SI{100}{\kilo\ohm} & *\SI{10}{\nano\farad} & 159.24    	& 1.14E-4		& 1.0005685 	& 24 & 5.6 \\ 
\SI{100}{\kilo\ohm} & *\SI{1}{\nano\farad} & 1592.36  	& 2.60E-4		& 1.0003486 	& 11 & 2.1 \\ 
\SI{10}{\kilo\ohm} & *\SI{10}{\nano\farad} & 1592.36  	& 1.34E-4		& 1.0007160 	& 24 & 3.5 \\ 
\SI{10}{\kilo\ohm} & *\SI{1}{\nano\farad} & 15873.02  	& 3.71E-4		& 0.9974559 	& 105 & 27 \\
\bottomrule
\end{tabular}
\label{tab:results}
\end{table*}

\begin{table*}[t]
	\centering
	\caption{Uncertainty budget for $W$, for $Z_\textup{A} = \SI{100}{\kilo\ohm}$, $Z_\textup{B} = \SI{1}{\nano\farad}$, $f=\SI{1592.36}{\hertz}$.}
	\label{tab:uncbug}
	\begin{tabular}{lllll}
		\addlinespace[5pt]
		\toprule
		Quantity & $X$ & $u(\varRe X)$ & $u(\varIm X)$ & type \\
		\midrule
		$\Delta g_\textup{FR}$ 				& \num{0+j0} 						& \num{e-6} 				& \num{e-6} & B\\
		$z_1, z_2$ 	 						& \SI{100+j40}{\milli\ohm} 			& \SI{50}{\milli\ohm}	& \SI{10}{\milli\ohm}  & B \\
		$y_\textup{HA},y_\textup{HB}$ 		& \SI{0+j2}{\micro\siemens}			& 0 						& \SI{0.5}{\micro\siemens} & B \\
		$W^{(\textup{r})}$					& $\num{2.610e-4}+\num{j1.0003500}$	& \num{E-7} 				& \num{E-7} & A \\	
		\midrule
		$W$									& $\num{2.604e-4}+\num{j1.0003486}$	& \num{6.3E-7}   		& \num{6.3e-7} &  \\
		\midrule
		$W^\textup{ref}$						& $\num{2.496e-4}+\num{j1.0003465}$	& \num{5.0E-6}   		& \num{5.8e-6} &  \\
		\bottomrule
	\end{tabular}
\end{table*}

\begin{figure}[htb]
	\centering
	\includegraphics[width=\columnwidth]{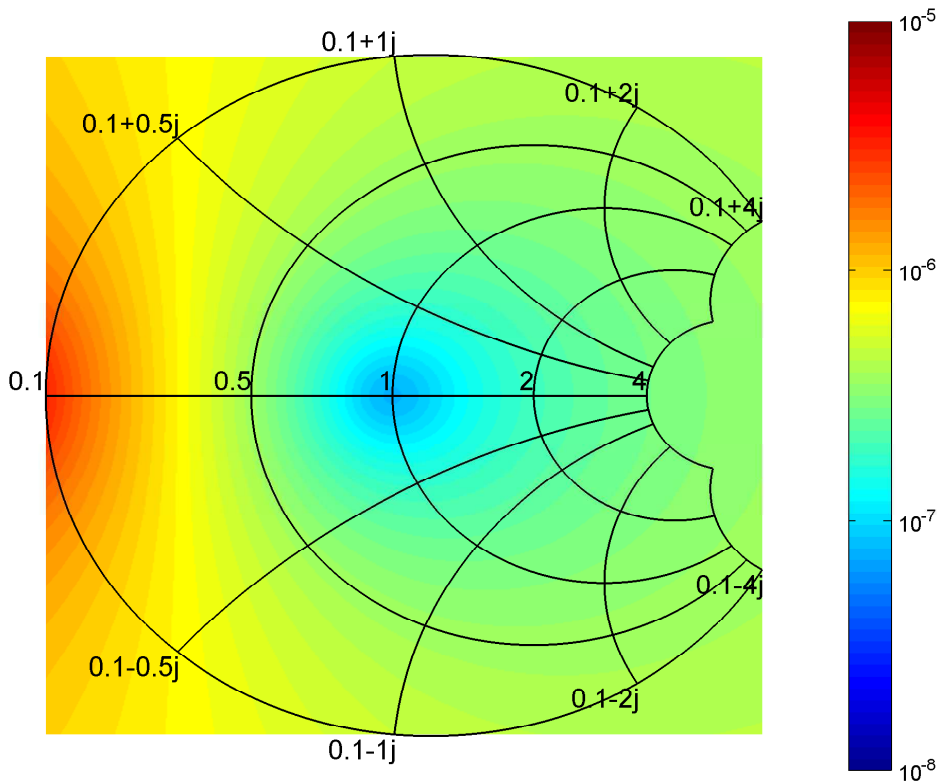}
	 \caption{Color plot of the magnitude of the relative uncertainty of $W$ for $Z_\textup{B} = \SI{100}{\kilo\ohm}$, and $z_k$ and $y_\textup{HX}$ as given in Tab.~\protect\ref{tab:uncbug}. The plot is given as Smith chart: $W$ coordinates are drawn as black lines; real values of $W$ are along the horizontal diameter; the plot center corresponds to $W = 1+\ju 0$ for which the best uncertainty is achieved.}
	\label{fig:uncmap}
\end{figure}

The uncertainty expression can be extend to arbitrary $W$ values provided that sufficient information about the input quantities is given. As an example, Fig.~\ref{fig:uncmap} shows a color plot of the magnitude $|u(W)|/|W|$ as a function of $W$, calculated for $Z_\textup{B} = \SI{100}{\kilo\ohm}$, $z_k$ and $y_\textup{HX}$ as given in Tab.~\ref{tab:uncbug}, and $|W|$ between $0.1$ and $10$; for convenience the plot is given as a Smith chart, that is, the cartesian coordinates correspond to the conformal mapping $(W-1)/(W+1)$. Since, at the moment, the characterization of $\Delta g_\textup{FR}$ is not complete, the plot does not take into account this specific contribution. Indeed, different values of $Z_\textup{B}$, $z_k$ and $y_\textup{HX}$ will lead to a different but analog plot. In particular, the uncertainty is expected to increase toward lower values of $|W|$ because, for fixed $Z_\textup{B}$, $Z_\textup{A}$ decreases.

\section{Conclusions}
The digital coaxial voltage ratio bridge realized allows to measure two terminal-pair impedances having arbitrary magnitude ratio and phase difference in the audio frequency range. The comparisons performed suggest a base accuracy in the \num{E-6} range with the commercial source currently employed. 

The development of this bridge is part of the European Metrology Research Programme (EMRP) Project SIB53 AIM~QuTE, \emph{Automated impedance metrology extending the quantum toolbox for electricity}. Deliverables of the project include the development of more accurate digital sources which will increase the accuracy of the bridge here described, and interlaboratory comparisons with special standards that will allow the validation of the bridge measurements.

\section*{Acknowledgment}
The authors are indebted with Jaani Nissil\"{a}, MIKES, Finland, for help in the set-up of the bridge.

The activity has been partially financed by Progetto Premiale MIUR\footnote{\emph{Ministero dell'Istruzione, dell'Universit\`a e della Ricerca.}}-INRIM P4-2011 \emph{Nanotecnologie per la metrologia elettromagnetica} and P6-2012 \emph{Implementation of the new International System (SI) of Units}. The work has been realized within the EMRP Project SIB53 AIM~QuTE, \emph{Automated impedance metrology extending the quantum toolbox for electricity}. The EMRP is jointly funded by the EMRP participating countries within EURAMET and the European Union. 




%

\bibliographystyle{IEEEtran}
\bibliography{DigitalImpedanceBridges}

\begin{thebibliography}{10}
\providecommand{\url}[1]{#1}
\csname url@samestyle\endcsname
\providecommand{\newblock}{\relax}
\providecommand{\bibinfo}[2]{#2}
\providecommand{\BIBentrySTDinterwordspacing}{\spaceskip=0pt\relax}
\providecommand{\BIBentryALTinterwordstretchfactor}{4}
\providecommand{\BIBentryALTinterwordspacing}{\spaceskip=\fontdimen2\font plus
\BIBentryALTinterwordstretchfactor\fontdimen3\font minus
  \fontdimen4\font\relax}
\providecommand{\BIBforeignlanguage}[2]{{%
\expandafter\ifx\csname l@#1\endcsname\relax
\typeout{** WARNING: IEEEtran.bst: No hyphenation pattern has been}%
\typeout{** loaded for the language `#1'. Using the pattern for}%
\typeout{** the default language instead.}%
\else
\language=\csname l@#1\endcsname
\fi
#2}}
\providecommand{\BIBdecl}{\relax}
\BIBdecl

\bibitem{Callegaro2014_CPEMDigitalBridge}
L.~Callegaro, V.~D'Elia, and F.~Pourdanesh, ``Experiences with a two
  terminal-pair digital impedance bridge,'' in \emph{Conf. Precision
  Electromagn. Meas. (CPEM)}, Rio de Janeiro, Brazil, 24--29 Aug 2014.

\bibitem{CallegaroBook}
L.~Callegaro, \emph{Electrical impedance: principles, measurement, and
  applications}, ser. in Sensors.\hskip 1em plus 0.5em minus 0.4em\relax Boca
  Raton, FL, USA: CRC press: Taylor \& Francis, 2013, iSBN: 978-1-43-984910-1.

\bibitem{Kim2012}
D.~B. Kim, K.-T. Kim, M.-S. Kim, K.~M. Yu, W.-S. Kim, and Y.~G. Kim,
  ``All-around dual source impedance bridge,'' in \emph{Conf. Precision
  Electromagn. Meas. (CPEM)}, Washington DC, USA, 1-6 Jul 2012, pp. 592--593.

\bibitem{Lan2012}
J.~Lan, Z.~Zhang, Z.~Li, Q.~He, J.~Zhao, and Z.~Lu, ``A digital compensation
  bridge for {$R$-$C$} comparisons,'' \emph{Metrologia}, vol.~49, no.~3, p.
  266, 2012.

\bibitem{Kampik:2014a}
M.~Kampik, ``Analysis of the effect of {DAC} resolution on {AC} voltage
  generated by digitally synthesized source,'' \emph{IEEE Trans. Instrum.
  Meas.}, vol.~63, pp. 1235--1243, 2014.

\bibitem{Callegaro2005}
L.~Callegaro, ``On strategies for automatic bridge balancing,'' \emph{{IEEE}
  Trans. Instr. Meas.}, vol.~54, no.~2, pp. 529--532, Apr 2005.

\bibitem{Kampik:2014b}
M.~Kampik and J.~Nissil\"{a}, ``Sib53 {AIM QuTE} visit report,'' MIKES, Tech.
  Rep., 2014.

\bibitem{Nissila2014}
J.~Nissil\"{a}, K.~Ojasalo, M.~Kampik, J.~Kaasalainen, V.~Maisi, M.~Casserly,
  F.~Overney, A.~Christensen, L.~Callegaro, V.~D'Elia, N.~T.~M. Tran,
  F.~Pourdanesh, M.~Ortolano, D.~B. Kim, J.~Penttil\"{a}, and L.~Roschier, ``A
  precise two-channel digitally synthesized {AC} voltage source for impedance
  metrology,'' in \emph{Conf. Precision Electromagn. Meas. (CPEM)}, Rio de
  Janeiro, Brazil, 24-29 Aug 2014.

\bibitem{Callegaro2005b}
L.~Callegaro, V.~D'Elia, and D.~Serazio, ``10-n{F} capacitance transfer
  standard,'' \emph{IEEE Trans. Instr. Meas.}, vol.~54, no.~5, pp. 1869--1872,
  2005.

\bibitem{GUMS2}
\BIBentryALTinterwordspacing
``{JCGM} 102:2011, {E}valuation of measurement data --- {S}upplement 2 to the
  ``{G}uide to the expression of uncertainty in measurement'' --- {E}xtension
  to any number of output quantities,'' 2011. [Online]. Available:
  \url{http://www.bipm.org}
\BIBentrySTDinterwordspacing

\bibitem{Zeier2012}
M.~Zeier, J.~Hoffmann, and M.~Wollensack, ``\texttt{Metas.UncLib} - a
  measurement uncertainty calculator for advanced problems,''
  \emph{Metrologia}, vol.~49, no.~6, p. 809, 2012.

\end{thebibliography}
\end{document}